\documentstyle[12pt]{article}

\title{Different approaches in the theory of the metastable phase decay on
several types of heterogeneous centers}

\author{V.Kurasov}

\date{Victor.Kurasov@pobox.spbu.ru}

\begin{document}

\maketitle


The problem of the metastable phase decay remains one of the most actual
questions  in the field of  kinetics of the first order phase transitions.
The first theories \cite{Wakeshima},  \cite{Kaban}  determined only the
characteristic time lag of the nucleation process.
The first well grounded theory  which allows to describe the evolution
of the system
during the nucleation period and to find the form of the droplets size
spectrum was presented in \cite{Kuni} where the iteration procedure was
formulated.

Now the theoretical description is going to be spread on the systems with
the several  types  of  heterogeneous  centers.  Unfortunately  the
direct generalization
of the iteration method  failed due to existence of the specific interaction
between
different sorts of heterogeneous centers through the vapor consumption.

The first attempt to solve
this problem was done in \cite{Multidec} where the structure of announced
description was composed by many asymptotic cases including
the intermediate case where the new special monodisperse approximation was
suggested.

This monodisperse approximation was well grounded but there was no
evident correspondence
between the iteration procedure and the
results presented in \cite{Multidec}.
Following  the  suggestion   of   F.M.Kuni   to   seek   for   this
correspondence we
shall present here the iteration solution of this problem.

We shall consider the physical situation from \cite{Multidec} and use
all definitions introduced there.

The structure of the publication will be the following:
\begin{itemize}

\item
In the first part the iteration procedure will be presented.

\item
In the second part we shall calculate iterations on the base of the property
of the "avalanche consumption" of the metastable phase.

\item
In the third part we shall calculate the iterations based on the special
monodisperse approximation.

\end{itemize}

\section{Iteration method}

We shall present all formulas for the case of two sorts of heterogeneous
centers. They will be marked by two subscripts $A$ and $B$.
The generalization for the arbitrary number of heterogeneous
centers types is evident. It is one of the real advantages of the presented
theory.

The system of the condensation equations  according to \cite{Multidec}
can be
presented as following
$$
G_A(z) = F_A \int_0^{z} (z-x)^3 \exp(\Gamma_A (\zeta(x) - \Phi)) \theta_A(x)
dx
$$
$$
G_B(z) = F_B \int_0^{z} (z-x)^3 \exp(\Gamma_B (\zeta(x) - \Phi)) \theta_B(x)
dx
$$
$$
\Phi = \zeta(z) +G_A(z) + G_B(z)
$$
$$
\theta_A(z) = \exp(-K_A \int_0^{z} \exp(\Gamma_A (\zeta(x) - \Phi))
dx           )
$$
$$
\theta_B(z) = \exp(-K_B \int_0^{z} \exp(\Gamma_B (\zeta(x) - \Phi))
dx           )
$$
with six parameters $F_A$, $F_B$, $K_A$, $K_B$, $\Gamma_A$,
$\Gamma_B$. Note  that
the definition of $\Gamma_A, \Gamma_B$ slightly differs from the standard
one - it is $\Phi$ times less.

The presented system can be directly seen from the condensation equations
system for one type of heterogeneous centers and there is no need to discuss
it more. The problem is how to solve it.

According to the scale invariance one can put $F_A = 1$, $\Gamma_A = 1$.
According
to the symmetry of the types one can choose them as to have $F_B < 1$.
These  simplifications  will  be  interesting  only  in   numerical
modeling
and here we shall keep all parameters.

We shall define the iteration procedure by the following relations
$$
G_{A i+1}(z) = F_A \int_0^{z} (z-x)^3 \exp(\Gamma_A (\zeta_{i}(x) - \Phi))
\theta_{Ai}(x)
dx
$$
$$
G_{Bi+1}(z) = F_B \int_0^{z} (z-x)^3 \exp(\Gamma_B (\zeta_{i}(x) - \Phi))
\theta_{Bi}(x)
dx
$$
$$
\Phi = \zeta_{i}(z) +G_{Ai}(z) + G_{Bi}(z)
$$
$$
\theta_{Ai+1}(z) = \exp(-K_A \int_0^{z} \exp(\Gamma_A (\zeta_{i}(x) - \Phi))
dx           )
$$
$$
\theta_{Bi+1}(z) = \exp(-K_B \int_0^{z} \exp(\Gamma_B (\zeta_{i}(x) - \Phi))
dx           )
$$
The number of the iteration approximation is marked by the ordinary subscript
instead of the capital letter which marks below the sort of the centers.
The values without the number of iteration will mark the real solutions.

The initial approximations are the following
$$
G_{A0} = 0, \ \ \ G_{B0} = 0 , \ \ \ \theta_{A0} = 1 , \ \ \ \theta_{B0} = 1
$$

Then one can note the following important chains

$$
G_{A0} < G_{A2} < ... < G_{A2i} < ... < G_A < ... < G_{A2i+1} < ... <
G_{A3} < G_{A1}
$$
$$
G_{B0} < G_{B2} < ... < G_{B2i} < ... < G_B < ... < G_{B2i+1} < ... <
G_{B3} < G_{B1}
$$
$$
\theta_{A0} > \theta_{A2} > ... > \theta_{A2i} > ... > \theta_A > ... >
\theta_{A2i+1} > ... >
\theta_{A3} > \theta_{A1}
$$
$$
\theta_{B0} > \theta_{B2} > ... > \theta_{B2i} > ... > \theta_B > ... >
\theta_{B2i+1} > ... >
\theta_{B3} > \theta_{B1}
$$
$$
\zeta_{0} > \zeta_{2} > ... > \zeta_{2i} > ...
> \zeta > ... > \zeta_{2i+1} > ... >
\zeta_{3} > \zeta_{1}
$$
for all values of arguments.

These chains prove the convergence of iterations and allow to estimate
the accuracy at every step of the calculation.

The problem is to calculate the iterations. In the situation with one
type of heterogeneous centers the system of condensation equations can
be gotten
 if we chancel $G_B$ and the index $A$. The same
thing we have to do with
the iteration procedure and one can get the mentioned properties.
In \cite{Kuni} only one iteration
for $G$ and two first iterations for $\theta$ were calculated. The further
iterations can not be calculated analytically. In the situation of
nucleation in the system with one type of heterogeneous centers it is
sufficient. But in the situation with several types of heterogeneous centers
it isn`t so. To show this we has to recall the physical sense of the first
iterations.

The first iteration for $G$ is calculated  on the base of
unexhausted number of heterogeneous centers.
So, the behavior of $\zeta$ is wrong when the power  of exhaustion is
essential. But when the centers are exhausted the number of droplets is
known - it is equal to the initial number of centers. This primitive notation
is the reason why the second iteration for $\theta$ gives already suitable
result.

In the situation with the several types of heterogeneous centers the situation
is different. Really, it is quite possible to have the exhaustion of
the first type of heterogeneous centers and the moderate exhaustion of
the second type centers. Then
the exhaustion  of the first type centers has to be
taken into account in calculation of the supersaturation
which is necessary to
get the number of the droplet appeared on the second type centers. But
as far as the power of exhaustion of the second type centers is moderate
one can not say that the number of the droplets on the second type centers
is equal to the total number of the second type centers. So, we have to
know the behavior of supersaturation in all cases and it is impossible
to do
already in the first iterations. Then we have to calculate further iterations.

Unfortunately it is impossible to calculate the further iteration
approximations
without any simplifications. That's why two ways of possible approximations
are presented in the next sections.

\section{Avalanche consumption of the metastable phase}

The sequential calculation of iterations gives
$$
G_{A1} = F_A \frac{z^4}{4}
$$
$$
\theta_{A1} = \exp(-K_A z)
$$
$$
G_{B1} = F_B \frac{z^4}{4}
$$
$$
\theta_{B1} = \exp(-K_B z)
$$
$$
G_{A2} = F_A \int_0^z (z-x)^3
\exp(-\Gamma_A( F_A + F_B) x^4 / 4)
\exp(-K_A x) dx
$$
$$
G_{B2} = F_B \int_0^z (z-x)^3
\exp(-\Gamma_B( F_A + F_B) x^4 / 4)
\exp(-K_B x) dx
$$


The problem is how to calculate the integrals for $G_{A2}, G_{B2}$. To
calculate these integrals
we shall extract functions
$$
\varphi_A = \exp(-\Gamma_A (G_A + G_B) )
$$
$$
\varphi_B = \exp(-\Gamma_B (G_A + G_B) )
$$
These functions can be calculated in the iteration approximation as
$$
\varphi_{Ai} = \exp(-\Gamma_A (G_{Ai} + G_{Bi}) )
$$
$$
\varphi_{Bi} = \exp(-\Gamma_B (G_{Ai} + G_{Bi}) )
$$

One can see that
$\varphi_{A1}, \varphi_{B1}$ can be well approximated as the step functions.
Really, $\exp(-x^4)$ can be approximately interpreted as the step
function\footnote{This
 was the reason why in experiments one can observe characteristic
time lag and this characteristic was introduced in the theory of the
metastable
phase decay. Then the first theoretical descriptions adopted this fact
as the given one without any justification.}.

Then
$$
G_{A2} = F_A \int_0^z (z-x)^3  \exp(-K_A x) \Theta(z_{1A} - x) dx
$$
$$
G_{B2} = F_B \int_0^z (z-x)^3  \exp(-K_B x) \Theta(z_{1B} - x) dx
$$
where
$$
z_{1A } \equiv  (\frac{4}{\Gamma_A (F_A + F_B) } )^{1/4}
$$
$$
z_{1B } \equiv  (\frac{4}{\Gamma_B (F_A + F_B) } )^{1/4}
$$
and these integrals can be easily calculated\footnote{More
correctly one gas to multiply $z_{1A}$ and $z_{1B}$ on $\ln^{1/4} 2$.
This corresponds to the definition of the halfwidth at the halfheight.}.

For $z>z_{1A}$
we have
$$
G_{A2} = \sum_{i=0}^{3}  p_{Ai} z^i
$$
where
$$
p_{Ai} = \frac{3!(-1)^i}{i! (3-i)!}
F_A \int_0^{z_{1A}} x^{3-i} \exp(-K_A x) dx
$$
are some constants which can be calculated analytically.

For $z>z_{1B}$
we have
$$
G_{B2} =
 \sum_{i=0}^{3}  p_{Bi} z^i
$$
where
$$
p_{Bi} = \frac{3!(-1)^i}{i! (3-i)!}
F_B \int_0^{z_{1B}} x^{3-i} \exp(-K_B x) dx
$$
are some constants calculated analytically.

For $z<z_{1A}$ we have
\begin{eqnarray}
G_{A2} =
F_A \exp(-K_A z) \frac{1}{K_A^4}
[ (z K_A)^3 \exp(zK_A) -
\nonumber
\\
\nonumber
3 (z K_A)^2 \exp(zK_A) +
6 z K_A \exp(zK_A) - 6 (\exp(K_A z) - 1) ]
\end{eqnarray}

For $z<z_{1B}$ we have
\begin{eqnarray}
G_{B2} =
F_B \exp(-K_B z) \frac{1}{K_B^4}
[ (z K_B)^3 \exp(zK_B) -
\nonumber
\\
\nonumber
3 (z K_B)^2 \exp(zK_B) +
6 z K_B \exp(zK_B) - 6 (\exp(K_B z) - 1) ]
\end{eqnarray}


The calculation of $\theta_{A2}$, $\theta_{A2}$ can be done in the same
way
$$
\theta_{A2} = \exp(- K_A \int_0^z
\exp(-\Gamma_A ( F_A + F_B) \frac{x^4}{4})  dx )
$$
$$
\theta_{B2} = \exp(- K_B \int_0^z
\exp(-\Gamma_B ( F_A + F_B) \frac{x^4}{4})  dx )
$$
Then
$$
\theta_{A2} = \exp(- K_A [ \Theta(z_{1A} - z) z + \Theta(z-z_{1A}) z_{1A} ] )
$$
$$
\theta_{B2} = \exp(- K_B [ \Theta(z_{1B} - z) z + \Theta(z-z_{1B}) z_{1B} ] )
$$

Now we have to calculate
$\theta_{A3}, \theta_{B3}$. One can note that functions
$\varphi_{A2}, \varphi_{B2}$ can be also presented as the step functions.
Namely, all functions
\begin{eqnarray}
\exp(
- \Gamma_A (
F_A \exp(-K_A z) \frac{1}{K_A^4}
[ (z K_A)^3 \exp(zK_A) -
\nonumber
\\
3 (z K_A)^2 \exp(zK_A) +
6 z K_A \exp(zK_A) - 6 (\exp(K_A z) - 1) ]
+
\nonumber
\\
\nonumber
F_B \exp(-K_B z) \frac{1}{K_B^4}
[ (z K_B)^3 \exp(zK_B) -
3 (z K_B)^2 \exp(zK_B) +
\\
\nonumber
6 z K_B \exp(zK_B) - 6 (\exp(K_B z) - 1) ]
)
)
\end{eqnarray}
\begin{eqnarray}
\exp(-
\Gamma_A (
F_A \exp(-K_A z) \frac{1}{K_A^4}
[ (z K_A)^3 \exp(zK_A) -
3 (z K_A)^2 \exp(zK_A) +
\nonumber
\\
\nonumber
6 z K_A \exp(zK_A) - 6 (\exp(K_A z) - 1) ]
+
 \sum_{i=0}^{3}  p_{Bi} z^i
)
)
\end{eqnarray}
$$
\exp(
- \Gamma_A (
 \sum_{i=0}^{3}  p_{Ai} z^i
+
 \sum_{i=0}^{3}  p_{Bi} z^i
)
)
$$
\begin{eqnarray}
\exp( -
\Gamma_A (
 \sum_{i=0}^{3}  p_{Ai} z^i
+
F_B \exp(-K_B z) \frac{1}{K_B^4}
[ (z K_B)^3 \exp(zK_B) -
\nonumber
\\
\nonumber
3 (z K_B)^2 \exp(zK_B) +
6 z K_B \exp(zK_B) - 6 (\exp(K_B z) - 1) ]
)
)
\end{eqnarray}
\begin{eqnarray}
\exp( -
\Gamma_B (
F_A \exp(-K_A z) \frac{1}{K_A^4}
[ (z K_A)^3 \exp(zK_A) -
\nonumber
\\
3 (z K_A)^2 \exp(zK_A) +
6 z K_A \exp(zK_A) - 6 (\exp(K_A z) - 1) ]
+
\nonumber
\\
\nonumber
F_B \exp(-K_B z) \frac{1}{K_B^4}
[ (z K_B)^3 \exp(zK_B) -
3 (z K_B)^2 \exp(zK_B) +
\\
\nonumber
6 z K_B \exp(zK_B) - 6 (\exp(K_B z) - 1) ]
)
)
\end{eqnarray}
\begin{eqnarray}
\exp( -
\Gamma_B (
F_A \exp(-K_A z) \frac{1}{K_A^4}
[ (z K_A)^3 \exp(zK_A) -
3 (z K_A)^2 \exp(zK_A) +
\nonumber
\\
\nonumber
6 z K_A \exp(zK_A) - 6 (\exp(K_A z ) - 1) ]
+
 \sum_{i=0}^{3}  p_{Bi} z^i
)
)
\end{eqnarray}
$$
\exp( -
\Gamma_B (
 \sum_{i=0}^{3}  p_{Ai} z^i
+
 \sum_{i=0}^{3}  p_{Bi} z^i
)
)
$$
\begin{eqnarray}
\exp( -
\Gamma_B (
 \sum_{i=0}^{3}  p_{Ai} z^i
+
F_B \exp(-K_B z) \frac{1}{K_B^4}
[ (z K_B)^3 \exp(zK_B) -
\nonumber
\\
\nonumber
3 (z K_B)^2 \exp(zK_B) +
6 z K_B \exp(zK_B) - 6 (\exp(K_B z) - 1) ]
)
)
\end{eqnarray}
have the step-like behavior. This is the central point of our calculations.
This property can be directly seen by calculations.

The possibility to see the step-like behavior of  exponent of the
supersaturation deviation directly on the base of explicit expressions
is the evident advantage of the iteration method.
Otherwise we have to prove this property in the general situation taking
into account that exponent of the supersaturation deviation lies
between $\exp(-z^4)$ and $\exp(-z^3)$ after the suitable renormalization.

Now to calculate $\theta_{A3}$, $\theta_{B3}$ we have to define
the values $z_{2A}$, $z_{2B}$ by equations
$$
\Gamma_A(G_{A2}(z_{2A}) + G_{B2}(z_{2A})) = 1
$$
$$
\Gamma_B(G_{A2}(z_{2B}) + G_{B2}(z_{2B})) = 1
$$
Then
$$
\theta_{A3} = \exp(- K_A [ \Theta(z_{2A} - z) z + \Theta(z-z_{2A}) z_{2A} ] )
$$
$$
\theta_{B3} = \exp(- K_B [ \Theta(z_{2B} - z) z + \Theta(z-z_{2B}) z_{2B} ] )
$$


The calculation of $G_{3A}$ and $G_{3B}$ can be done by the following
way
\begin{eqnarray}
G_{A3} = F_A \int_0^z
(z-x)^3
\exp(-\Gamma_A (G_{A2} + G_{B2}) )
\nonumber
\\
\nonumber
 \exp(- K_A [ \Theta(z_{1A} - x) x + \Theta(x-z_{1A}) z_{1A} ] )
dx
\end{eqnarray}
\begin{eqnarray}
G_{A3} =  F_A \int_0^z
(z-x)^3
\Theta(z_{2A} - x)
\nonumber
\\
\nonumber
 \exp(- K_A [ \Theta(z_{1A} - x) x + \Theta(x-z_{1A}) z_{1A} ] )
dx
\end{eqnarray}
\begin{eqnarray}
G_{B3} =  F_B \int_0^z
(z-x)^3
\exp(-\Gamma_B (G_{A2} + G_{B2}) )
\nonumber
\\
\nonumber
\exp(- K_B [ \Theta(z_{1B} - x) x + \Theta(x-z_{1B}) z_{1B} ] )
dx
\end{eqnarray}
\begin{eqnarray}
G_{B3} =  F_B \int_0^z
(z-x)^3
\Theta(z_{2B} - x)
\nonumber
\\
\nonumber
 \exp(- K_B [ \Theta(z_{1B} - x) x + \Theta(x-z_{1B}) z_{1B} ] )
dx
\end{eqnarray}

These integrals can be easily taken in analytical form.
For $G_{A3}$ one can get
\begin{itemize}
\item
For $z > z_{2A}$ with arbitrary $z_{1A}$ we have
$$
G_{A3} =
\sum_{i=0}^{3} z^i p_{Ai}
$$
where
\begin{eqnarray}
p_{Ai} = \frac{(-1)^i 3!}{i! (3-i)!} F_A
\int_0^{z_{2A}}
x^i [\exp(-K_A z_{1A}) \Theta(x-z_{1A}) +
\nonumber
\\
\nonumber
\exp(-K_A x) \Theta(z_{1A} - x) ] dx
\end{eqnarray}
can be calculated analytically.

\item
For $z< min \{ z_{1A} , z_{2A} \}$
$$
G_{A3} = F_A \int_0^z (z-x)^3 \exp(-K_A x) dx
$$
has been already calculated.

\item
For $z_{1A} < z < z_{2A}$
$$
G_{A3} = \sum_{i=0}^3 z^i p_{Ai} +
F_A \frac{(z-z_{1A})^4 }{4} \exp(-K_A z_{1A})
$$
where
$$
p_{Ai} = F_A
\frac{3! (-1)^i} {i! (3-i)!} \int_0^{z_{1A}} x^i \exp(-K_A x) dx
$$

\end{itemize}
All $p_{Ai}$ are constants and can be calculated analytically.

For $G_{B3}$ one can get
\begin{itemize}
\item
For $z > z_{2B}$ with arbitrary $z_{1B}$
$$
G_{B3} =
\sum_{i=0}^{3} z^i p_{Bi}
$$
where
\begin{eqnarray}
p_{Bi} = \frac{(-1)^i 3!}{i! (3-i)!} F_B
\int_0^{z_{2B}}
x^i
[\exp(-K_B z_{1B}) \Theta(x-z_{1B}) +
\nonumber
\\
\nonumber
\exp(-K_B x) \Theta(z_{1B} - x) ] dx
\end{eqnarray}

\item
For $z< min \{ z_{1B} , z_{2B} \}$
$$
G_{B3} = F_B \int_0^z (z-x)^3 \exp(-K_B x) dx
$$
has been already calculated.

\item
For $z_{1B} < z < z_{2B}$
$$
G_{B3} = \sum_{i=0}^3 z^i p_{Bi} +
F_B \frac{(z-z_{1B})^4 }{4} \exp(-K_B z_{1B})
$$
where
$$
p_{Bi} = F_B
\frac{3! (-1)^i} {i! (3-i)!} \int_0^{z_{1B}} x^i \exp(-K_B x) dx
$$

\end{itemize}
All $p_{Bi}$ are constants and can be calculated analytically.

One can easily see that
$z_{1A} < z_{2A}, \ \ z_{1B} < z_{2B}$ and the third opportunity can
take place.

Moreover, if we define
$
z_{iA}$, $z_{iB}$
by equalities
$$
\Gamma_A (G_{Ai} (z_{iA}) + G_{Bi} (z_{iA}) ) = 1
$$
$$
\Gamma_B (G_{Ai} (z_{iB}) + G_{Bi} (z_{iB}) ) = 1
$$
then we can
get we following chains of inequalities
$$
z_{1A} < z_{3A} < ... < z_{2i+1 A} < ... < z_{A} < ... < z_{2iA} < ...
< z_{4A} < z_{2A}
$$
$$
z_{1B} < z_{3B} < ... < z_{2i+1 B} < ... < z_{B} < ... < z_{2iB} < ...
< z_{4B} < z_{2B}
$$
where
$z_A$, $z_B$  are defined by
$$
\Gamma_A (G_{A} (z_{A}) + G_{B} (z_{A}) ) = 1
$$
$$
\Gamma_B (G_{A} (z_{B}) + G_{B} (z_{B}) ) = 1
$$


Then we can calculate $G_{A4}$, $G_{B4}$.
In the  initial expressions
$$
G_{A4} =
F_A \int_0^z (z-x)^3
\exp(-\Gamma_A (G_{A3} + G_{B3}) \theta_{A3} dx
$$
$$
G_{B4} =
F_B \int_0^z (z-x)^3
\exp(-\Gamma_A (G_{A3} + G_{B3}) \theta_{B3}  dx
$$ we have to use two characteristic lengths
$z_2$ and $z_3$.
Then
\begin{eqnarray}
G_{A4} =
F_A
\int_0^z
\Theta(z_{3A} - x)
(z-x)^3
\nonumber
\\
\nonumber
[ \exp(-K_A x) \Theta(z_{2A} - x) +
\exp(-K_A z_{2A}) \Theta(x-z_{2A}) ] dx
\end{eqnarray}
\begin{eqnarray}
G_{B4} =
F_B
\int_0^z
\Theta(z_{3B} - x)
(z-x)^3
\nonumber
\\
\nonumber
[ \exp(-K_B x) \Theta(z_{2B} - x) +
\exp(-K_B z_{2B}) \Theta(x-z_{2B}) ] dx
\end{eqnarray}

These integrals can be easily taken in an analytical form.

For $G_{A4}$ one can get
\begin{itemize}
\item
For $z > z_{3A}$ with arbitrary $z_{2A}$
$$
G_{A4} =
\sum_{i=0}^{3} z^i p_{Ai}
$$
where
\begin{eqnarray}
p_{Ai} = \frac{(-1)^i 3!}{i! (3-i)!} F_A
\int_0^{z_{3A}}
x^i [\exp(-K_A z_{2A}) \Theta(x-z_{2A}) +
\nonumber
\\
\nonumber
\exp(-K_A x) \Theta(z_{2A} - x) ] dx
\end{eqnarray}

\item
For $z< min \{ z_{2A} , z_{3A} \}$
$$
G_{A4} = F_A \int_0^z (z-x)^3 \exp(-K_A x) dx
$$
has been already calculated.

\item
For $z_{2A} < z < z_{3A}$
$$
G_{A4} = \sum_{i=0}^3 z^i p_{Ai} +
F_A \frac{(z-z_{2A})^4 }{4} \exp(-K_A z_{2A})
$$
where
$$
p_{Ai} = F_A
\frac{3! (-1)^i} {i! (3-i)!} \int_0^{z_{2A}} x^i \exp(-K_A x) dx
$$

\end{itemize}
All $p_{Ai}$ are constants and can be calculated analytically.

For $G_{B4}$ one can get
\begin{itemize}
\item
For $z > z_{3B}$ with arbitrary $z_{2B}$
$$
G_{B4} =
\sum_{i=0}^{3} z^i p_{Bi}
$$
where
\begin{eqnarray}
p_{Bi} = \frac{(-1)^i 3!}{i! (3-i)!} F_B
\int_0^{z_{3B}}
x^i [\exp(-K_B z_{2B}) \Theta(x-z_{2B}) +
\nonumber
\\
\nonumber
\exp(-K_B x) \Theta(z_{2B} - x) ] dx
\end{eqnarray}

\item
For $z< min \{ z_{2B} , z_{3B} \}$
$$
G_{B4} = F_B \int_0^z (z-x)^3 \exp(-K_B x) dx
$$
has been already calculated.

\item
For $z_{2B} < z < z_{3B}$
$$
G_{B4} = \sum_{i=0}^3 z^i p_{Bi} +
F_B \frac{(z-z_{2B})^4 }{4} \exp(-K_B z_{2B})
$$
where
$$
p_{Bi} = F_B
\frac{3! (-1)^i} {i! (3-i)!} \int_0^{z_{2B}} x^i \exp(-K_B x) dx
$$

\end{itemize}
All $p_{Bi}$ are constants and can be calculated analytically.

One can easily see that
$z_{3A} < z_{2A}, \ \ z_{3B} < z_{2B}$ and the third opportunity can not
take place.


One can calculate by the same way all further iterations\footnote{That's
why we left all possibilities in the last expressions.}.

We see that due to the difference of the relations between parameters
$z_{Ai}$, $z_{Bi}$ even the functional form of the iterations will be
different. But it is quite easy to see that calculations of further iterations
will be done absolutely analogously. Moreover, the functional forms of
all odd further iterations will be similar. The functional forms of all
even iterations will be also similar.

Then one can extract these functional forms and it is necessary to obtain
the values of parameters $z_{Ai}$, $z_{Bi}$ in these forms.

Moreover, one can put one and the same values of "final" parameters $z_A$,
$z_B$ in equations for $G_A$, $G_B$, $\theta_A$, $\theta_B$.  This leads
to the following result

In the  initial expressions
$$
G_{A} =
F_A \int_0^z (z-x)^3
\exp(-\Gamma_A (G_{A} + G_{B})) \theta_{A} dx
$$
$$
G_{B} =
F_B \int_0^z (z-x)^3
\exp(-\Gamma_A (G_{A} + G_{B})) \theta_{B}  dx
$$
Then
$$
G_{A} =
F_A
\int_0^z
\Theta(z_{A} - x)
(z-x)^3
 \exp(-K_A x)
dx
$$
$$
G_{B} =
F_B
\int_0^z
\Theta(z_{B} - x)
(z-x)^3
 \exp(-K_B x)  dx
$$

These integrals can be easily taken in analytical form.

For $G_{A}$ one can get
\begin{itemize}
\item
For $z > z_{A}$
$$
G_{A} =
\sum_{i=0}^{3} z^i p_{Ai}
$$
where
$$
p_{Ai} = \frac{(-1)^i 3!}{i! (3-i)!} F_A
\int_0^{z_{A}}
x^i
\exp(-K_A x) dx
$$

\item
For $z<  z_{A} $
$$
G_{A4} = F_A \int_0^z (z-x)^3 \exp(-K_A x) dx
$$
has been already calculated.

\end{itemize}
All $p_{Ai}$ are constants and can be calculated analytically.

For $G_{B}$ one can get
\begin{itemize}
\item
For $z > z_{B}$
$$
G_{B} =
\sum_{i=0}^{3} z^i p_{Bi}
$$
where
$$
p_{Bi} = \frac{(-1)^i 3!}{i! (3-i)!} F_B
\int_0^{z_{B}}
x^i
\exp(-K_B x) dx
$$

\item
For $z<  z_{B} $
$$
G_{B4} = F_B \int_0^z (z-x)^3 \exp(-K_B x) dx
$$
has been already calculated.

\end{itemize}
All $p_{Bi}$ are constants and can be calculated analytically.

As the result now we know the functional form for $G_A$, $G_B$, $\theta_A$,
$\theta_B$. The condensations equations system now determines the
characteristic
values $z_{A}$, $z_{B}$.

The step-like behavior of $\varphi$ occurs due to the avalanche character
of the vapor consumption. The avalanche character  of the vapor consumption
has to be considered as the main feature of the first order phase transition
kinetics. This feature was described in \cite{book1} in details. Meanwhile
it is possible to make the theory more accurate taking into account the
regime of the droplets growth.

Note that all functions $\varphi$ are lying between $\exp(-x^4)$ and
$\exp(-x^3)$
after the suitable renormalization. Namely, they are more sharp than
$\exp(-x^3)$
and more smooth than $\exp(-x^4)$. Note that
$$
\int_0^{\infty} \exp(-x^4) dx \approx \int_0^{\infty}
\exp(-x^3) dx \approx 0.9 \equiv
D
$$
Then one can make results of the theory more accurate if we take
\begin{itemize}
\item
instead
$\Theta (z_{iA} -x )$ the function $\Theta ( D z_{iA} - x)$
\item
instead
$\Theta (z_{iB} -x )$ the function $\Theta ( D z_{iB} - x)$
\item
instead
$\Theta ( x- z_{iA}  )$ the function $ D \Theta ( x - D z_{iA} )$
\item
instead
$\Theta ( x- z_{iB}  )$ the function $ D \Theta ( x - D z_{iB} )$
\end{itemize}
and  multiply all parameters $z_{iA}$, $z_{iB}$ on $D$.

We can see that the generalization of the presented approach to the
multicomponent
case is quite evident. But here the number of unknown parameters $z_{A}$
will be equal to the number of the sorts of heterogeneous centers. The
problem how to solve this system is rather actual and here we shall propose
a way to do it.

We shall choose the sorts of heterogeneous centers as to have
$$
\Gamma_A > \Gamma_B > \Gamma_C > \Gamma_D >    ....
$$
Then
$$
z_{A} < z_{B} < z_{C} < z_{D} <  ......
$$

At first we shall determine $z_{A}$. Note that the behavior of the system
after $z_{A}$. Then we can put all
all $\Gamma_{B}$, $\Gamma_{C}$ etc. to be equal to $\Gamma_{A}$ and
$z_{B}$, $z_{C}$ etc. to be equal $z_{A}$.
Then we have only one parameter $z_{A}$ and can determine it from one
algebraic equation analogous to those described in \cite{book2} (the
situation is absolutely analogous to the intermediate situation).

Now $z_{A}$ is determined and $G_{A}$, $\theta_{A}$ are known. It allows
to go to the determination of $z_{B}$. It  can be done by the same procedure.
We have to consider $G_{A}$, $\theta_{A}$ as known values and put all
$z_{C}$, $z_{D}$ etc. to be equal to $z_{B}$. Then the analogous equation
can be solved and it gives $z_{B}$, $G_{B}$, $\theta_{B}$. This procedure
can be continued. Note that at first this way was described in \cite{book2}
for the special monodisperse approximation.

\section{Special monodisperse approximation}

The task to give the theoretical description of the complex systems required
 new approximations for some characteristic functions. As the result the
monodisperse approximation for $G_A$ and $G_B$ was suggested in
\cite{Multidec}.
The monodisperse approximation can be used in two variants - the first
one is the monodisperse approximation
with fixed number of droplets which is suitable when $G_A$, $G_B$
are really important, the second one is the monodisperse approximation
with a floating number of droplets involved in this approximation. A special
recipe allows to get approximation which is suitable during all times.

The second variant of approximation is necessary for the systems with
the strong hierarchy between the probabilities  of droplets formation on
different sorts of centers. Nevertheless in solution of the system by the
procedure analogous to the already described in the end of the
last section it is
possible to use only the first type of approximation. This states the
significance of the first type of approximation. But as far as we follow
here the iteration procedure it will be necessary to use the second variant
of approximation.

The problems to calculate iterations appear in the second approximation:
$$
G_{A2} = F_A \int_0^z (z-x)^3
\exp( -\Gamma_A (F_A + F_B) \frac{x^4}{4})
\exp(-K_A x) dx
$$
$$
G_{B2} = F_B \int_0^z (z-x)^3  \exp( -\Gamma_B (F_A + F_B) \frac{x^4}{4}
)
\exp(-K_B x) dx
$$

The justification of the monodisperse approximation has been already presented
many times (see \cite{Multidec}, \cite{book1}) and we
shall use it here without discussion. Then
$$
G_{A2} \approx
F_A z^3 \int_0^{z/4}    \exp( -\Gamma_A (F_A + F_B) \frac{x^4}{4}
)
\exp(-K_A x) dx
$$
$$
G_{B2} \approx
F_B z^3 \int_0^{z/4}    \exp( -\Gamma_B (F_A + F_B) \frac{x^4}{4}
)
\exp(-K_B x) dx
$$
The real advantage of consideration of the monodisperse approximation
on the level of iterations is the possibility to calculate the error of
approximation explicitly. This error  is small.

One can make the iteration approximation more accurate
by the substitution
of
 $\exp(-K_A x)$ by
$$\theta_{2A} = \exp(-K_A \int_0^z \exp(-\Gamma_A(F_A+ F_B) \frac{x^4}{4})
dx )
$$
and by the substitution
of
 $\exp(-K_B x)$ by
$$\theta_{2B} = \exp(-K_B \int_0^z \exp(-\Gamma_B(F_A+ F_B) \frac{x^4}{4})
dx )
$$

This leads to
$$
G_{2A} (z) = F_A ( 1  - \theta_{2A}(\frac{z}{4})
) z^3 / K_A $$
$$
G_{2B} (z) = F_B ( 1  - \theta_{2B}(\frac{z}{4})
) z^3 / K_B
$$

When monodisperse approximation is used it convenient to use the following
way to construct iterations
$$
G_{A i+1}(z) = F_A \int_0^{z} (z-x)^3 \exp(\Gamma_A (\zeta_{i}(x) - \Phi))
\theta_{Ai+1}(x)
dx
$$
$$
G_{Bi+1}(z) = F_B \int_0^{z} (z-x)^3 \exp(\Gamma_B (\zeta_{i}(x) - \Phi))
\theta_{Bi+1}(x)
dx
$$
All other formulas remain the same. Certainly the chains of inequalities
will be slightly violated, but this can not leads to divergence of iterations.
Moreover, one can simply add the difference between sequential $\theta$
to the error between sequential iterations.

Now we see that
 in monodisperse approximation
$$
G_{iA} (z) = \frac{ F_A z^3}{K_A}  ( 1 - \theta_{Ai}(z/4) )
$$
$$
G_{iB} (z) = \frac{F_B z^3}{K_B} ( 1 - \theta_{Bi}(z/4))
$$
So, we need to calculate only $\theta_{Ai}$, $\theta_{Bi}$.

In this calculation one can use the avalanche character of the vapor
consumption
again. Then
$$
\theta_{2A} (z)  = \exp(-K_A z_{1A} ) \theta (z-z_{1A}) +
\exp(-K_A z) \Theta(z_{1A} - z)
$$
$$
\theta_{2B} (z) = \exp(-K_B z_{1B} ) \theta (z-z_{1B}) +
\exp(-K_B z) \Theta(z_{1B} - z)
$$
where
$z_{1A}$, $z_{1B}$ are given by the previous expressions (but further
  $z_{iA}$, $z_{iB}$ will have slightly another numerical values).

Note that in practice we don't use such expression but can use the simple
asymptotes      ("essential asymptotes")
$$
\theta_{2A} =
\exp(-K_A z)
$$
$$
\theta_{2B} =
\exp(-K_B z)
$$
for all $z$.

Really, we need these expressions only for $z \leq z_{A}$,  $z \leq z_{B}$.
As far as the argument is $z/4$ it means that these asymptotes can not
be used only when
$z_{A} \geq 4 z_{B}$ or $z_{B} \geq  4 z_{A}$. It can take place only
when $\Gamma_A$ and $\Gamma_B$ differ more than in $4^3 = 64$ times.
The last seems to be practically unrealizable.

The third iteration for $\theta$ gives
\begin{eqnarray}
\theta_{3A} =
\exp(    -
K_A
\int_0^z
\exp(
-\Gamma_A
(
F_A
(
1 - \exp(-K_A \frac{x}{4})
\nonumber
\\
 \Theta(- \frac{x}{4} + z_{1A} )  -
 \exp(-K_A z_{1A}) \Theta( - z_{1A} + \frac{x}{4}  )
) +
\nonumber
\\
\nonumber
F_B
(
1 - \exp(-K_B \frac{x}{4}) \Theta( - \frac{x}{4} + z_{1B} )  -
\\
\nonumber
 \exp(-K_B z_{1B}) \Theta( - z_{1B} + \frac{x}{4}  )
)
)
)
dx
)
\end{eqnarray}
\begin{eqnarray}
\theta_{3B} =
\exp(    -
K_B
\int_0^z
\exp(
-\Gamma_B
(
F_A
(
1 - \exp(-K_A \frac{x}{4})
\nonumber
\\
\Theta(- \frac{x}{4} + z_{1A} )  -
 \exp(-K_A z_{1A}) \Theta(- z_{1A} + \frac{x}{4}  )
) +
\nonumber
\\ \nonumber
F_B
(
1 - \exp(-K_B \frac{x}{4}) \Theta(- \frac{x}{4} + z_{1B} )  -
\\
\nonumber
 \exp(-K_B z_{1B}) \Theta( - z_{1B} + \frac{x}{4}  )
)
)
)
dx
)
\end{eqnarray}

One can directly see that
\begin{eqnarray}
G_{A i+1} (z)  =
F_A z^3 \int_0^{z/4}
\exp( -\Gamma_A (G_{Ai} (x) + G_{Bi} (x) )    )
\nonumber
\\
\exp(- K_A \int_0^x
\exp(- \Gamma_A (G_{iA} (x') + G_{iB} (x')) ) dx' ) dx =
\nonumber \\
\nonumber
 \frac{F_A z^3}{K_A} (1-
\exp(- K_A \int_0^z
\exp(- \Gamma_A (G_{iA} (x) + G_{iB} (x)) ) dx ) )
\end{eqnarray}
\begin{eqnarray}
G_{B i+1} (z) =
F_B z^3  \int_0^{z/4}
\exp( -\Gamma_B (G_{Ai} (x) + G_{Bi} (x) )    )
\nonumber
\\
\exp(- K_B \int_0^x
\exp(- \Gamma_B (G_{iA} (x') + G_{iB} (x')) ) dx' ) dx =
\nonumber
\\ \nonumber
 \frac{F_B z^3}{K_B} (1-
\exp(- K_B \int_0^z
\exp(- \Gamma_B (G_{iA} (x) + G_{iB} (x)) ) dx ) )
\end{eqnarray}
Then $G_{A3}$, $G_{B3}$ have the same structure as $G_{A2}$, $G_{B2}$
have.
Certainly the numerical values of parameters $z_{Ai}$, $z_{Bi}$  can be
slightly another.
All  other iteration approximations will have the same form.

Now we can reformulate the system of the condensation equations because
we know the functional form of $G_{A}$, $G_{B}$. It will be the following
$$
G_A(z) = \frac{F_A z^3}{K_A} (1- \exp(-K_A \int_0^{z/4} \exp( - \Gamma_A
(G_{A}(x)
+ G_{B}(x) ))
dx           ) )
$$
$$
G_B(z) = \frac{F_B z^3}{K_B} (1- \exp(-K_B \int_0^{z/4} \exp( -
\Gamma_B (G_{A}
(x)
+ G_{B} (x) ))
dx           ) )
$$

In the avalanche approximation it will be the following
$$
G_A(z) = \frac{F_A z^3}{K_A} (1- \exp(-K_A z/4) \Theta (z_{A}  - z/4)
-
\exp(-K_A z_{A} \Theta(z/4 - z_{A})         )
$$
$$
G_B(z) = \frac{F_B z^3}{K_B} (1- \exp(-K_B z/4) \Theta (z_{B}  - z/4)
-
\exp(-K_B z_{B} \Theta(z/4 - z_{B})         )
$$
where
$z_{A}$, $z_{B}$ are defined by the previous relations.

In the approximation of essential asymptotes
$$
G_A(z) = \frac{F_A z^3}{K_A} (1- \exp(-K_A z/4)  )
$$
$$
G_B(z) = \frac{F_B z^3}{K_B} (1- \exp(-K_B z/4)   )
$$

These algebraic equations can be easily solved.

Now one can mention some ways to go away from the avalanche approximation.
In the approximation of essential asymptotes everything is clear. In the
avalanche approximation one can consider the number of the free heterogeneous
centers as the smooth function of $z$ and approximately consider
$$
G_A(z) = \frac{F_A z^3}{K_A} (1- \exp(-K_A \int_0^{z_A/4} \exp(\Gamma_A
(G_{A}(x)
+ G_{B}(x) ))
dx           ) )
$$
$$
G_B(z) = \frac{F_B z^3}{K_A} (1- \exp(-K_B \int_0^{z_B/4} \exp(\Gamma_B (G_{A}
(x)
+ G_{B} (x) ))
dx           ) )
$$

Then $G_A \sim const z^3$, $G_B \sim const z^3$, we have the monodisperse
approximation with fixed number of droplets and need only to calculate
the integrals of the type
$
\int_0^z \exp(-x^3) dx
$. This leads to the necessity to take
\begin{itemize}
\item
instead
$\Theta (z_{iA} -x )$ the function $\Theta ( D z_{iA} - x)$
\item
instead
$\Theta (z_{iB} -x )$ the function $\Theta ( D z_{iB} - x)$
\item
instead
$\Theta ( x- z_{iA}  )$ the function $ D \Theta ( x - D z_{iA} )$
\item
instead
$\Theta ( x- z_{iB}  )$ the function $ D \Theta ( x - D z_{iB} )$
\end{itemize}
and to  multiply all parameters $z_{iA}$, $z_{iB}$ on $D$. Here
$$
D \equiv \int_0^{\infty} \exp(-x^3)  dx = 0.9
$$

The generalization of the presented approach to the multicomponent
case is quite evident.
Our actions are absolutely identical to those described for the avalanche
model.
The only thing to stress is that at every step we can use
"essential asymptote"
for all sorts of heterogeneous centers whose $G_{A}$ are still undetermined.
Those $G_A$ which are already determined have to be taken explicitly.
 Note that at first this way was described in \cite{book2}.

As  the result we have to conclude that approximations of the avalanche
consumption
or the special monodisperse approximation lead to the similarity of the
iterations of the high order (at least in their functional form). So,
we see now the correspondence between the iteration approach and
approximations
used here. The numerical calculations show that the errors of these
approximations are negligible in frames of the accuracy of the modern
experiment.

\end{document}